\documentclass[%
 reprint,
 amsmath,amssymb,
 prd,
]{revtex4-1}

\usepackage{graphicx}
\usepackage{dcolumn}
\usepackage{bm}

\PassOptionsToPackage{breaklinks}{hyperref}
\usepackage{xurl} 
\usepackage{hyperref}
\usepackage{amsmath, amssymb, amsthm, mathtools, relsize, multirow, enumitem, dsfont}
\usepackage{comment}
\usepackage{simpler-wick}
\usepackage{amsmath}
\usepackage{graphicx,float}
\usepackage{appendix}
\usepackage{physics}
\usepackage{xcolor}
\usepackage{simpler-wick}
\usepackage{tikz}

\DeclareFontFamily{OMS}{oasy}{\skewchar\font48 }
\DeclareFontShape{OMS}{oasy}{m}{n}{%
         <-5.5> oasy5     <5.5-6.5> oasy6
      <6.5-7.5> oasy7     <7.5-8.5> oasy8
      <8.5-9.5> oasy9     <9.5->  oasy10
      }{}
\DeclareFontShape{OMS}{oasy}{b}{n}{%
       <-6> oabsy5
      <6-8> oabsy7
      <8->  oabsy10
      }{}
\DeclareSymbolFont{oasy}{OMS}{oasy}{m}{n}
\SetSymbolFont{oasy}{bold}{OMS}{oasy}{b}{n}
\DeclareMathSymbol{\smallleftarrow}     {\mathrel}{oasy}{"20}
\DeclareMathSymbol{\smallrightarrow}    {\mathrel}{oasy}{"21}
\DeclareMathSymbol{\smallleftrightarrow}{\mathrel}{oasy}{"24}

\newcommand{\ads}{AdS_5\times S^5}

\newcommand{\str}{\mathrm{sTr}}
\newcommand{\res}{\mathrm{Res}}

\newcommand{\Lax}{\mathcal{L}}

\begin{document}

\title{ 4D Chern-Simons and the pure spinor $\ads$ superstring}
\author{Nathan Berkovits}
 \email{nathan.berkovits@unesp.br}
\author{Rodrigo S. Pitombo}%
 \email{rs.pitombo@unesp.br}
\affiliation{ICTP South American Institute for Fundamental Research\\
Instituto de Física Teórica, UNESP - Univ. Estadual Paulista\\
Rua Dr. Bento Teobaldo Ferraz 271, 01140-070, São Paulo, SP, Brazil
}%


\begin{abstract}
   Four-dimensional Chern-Simons (4DCS) theory is useful for understanding integrable sigma-models and constructing new ones. In this paper, we show how to derive the complete pure spinor $\ads$ superstring sigma-model from 4DCS theory with defects. The matter sector of this sigma model was previously derived by Costello and Yamazaki, and we propose here that the pure spinor ghosts come from gauge-fixing meromorphic transformations of 4DCS which lead to the usual pure spinor Lax connection including the ghost contribution.
\end{abstract}

\maketitle
\tikzset{every picture/.style={line width=0.75pt}}
\section{Introduction}
An important step in fully understanding the AdS/CFT duality in its original formulation \cite{maldacena_large-n_1999} is to study the behavior of superstrings propagating in the $\ads$ background. In this context, the pure spinor formalism for the superstring \cite{berkovits_super-poincare_2000} plays an essential role since it can describe Type IIB backgrounds with Ramond-Ramond flux \cite{berkovits_ten-dimensional_2002} and its worldsheet action in an $\ads$ background is quantizable \cite{berkovits_quantum_2005}. 
Sigma models that describe superstrings in this background are special because they are integrable (at least classically). This means that the equations of motion for these models are equivalent to the flatness of a Lax connection $\Lax(z)$: 
\begin{equation}
    d\Lax(u)+\Lax(u)\wedge\Lax(u)=0\;\;\;\;\forall u \in \mathds{C} \label{LaxFlat}
\end{equation}
where $z$ is called the spectral parameter. The Green-Schwarz \cite{bena_hidden_2004}, pure spinor \cite{Brenno_Carlini_Vallilo_2004, berkovits_brst_2005} and the recently introduced B-RNS-GSS \cite{chandia_b-rns-gss_2023} formalisms have this property. Moreover, integrable structures also appear in the field theory side of the duality \cite{minahan_bethe-ansatz_2003, beisert_n4_2003}.

Discovering that a model has a Lax connection satisfying (\ref{LaxFlat}) usually involves guesswork. However, in a series of papers \cite{costello_gauge_2018-1,costello_gauge_2018,costello_gauge_2019} Costello, Witten and Yamazaki have shown that one can use a four-dimensional version of Chern-Simons theory with defects to construct sigma models in such a way that their integrability is guaranteed, as the flatness equation is implied by one of the equations of motion. For our purposes, the theory lives in $\mathds{R}^2\times \mathds{CP}^1$ and the spectral parameter on which $\Lax$ depends is reinterpreted as a holomorphic coordinate on the Riemann sphere.

Using 4D Chern-Simons theory (4DCS), several integrable field theories have been shown to come from different defect configurations in the 4D manifold such as the principal chiral model, the WZW model, the Yang-Baxter model \cite{costello_gauge_2019, lacroix_four-dimensional_2022}, and others \cite{fukushima_non-abelian_2022, fukushima_faddeev-reshetikhin_2021, liniado_integrable_2023,caudrelier_zakharovmikhailov_2021,lacroix_integrable_2021,vicedo_4d_2021}. Thus, a natural question is whether one can construct the superstring sigma model on $\ads$ with this framework. In \cite{costello_gauge_2019} Costello and Yamazaki constructed the matter sector (\ref{PS_Matter}) of the pure spinor superstring on $\ads$, and some integrable deformations \cite{fukushima_yang-baxter_2020,tian_ensuremathlambda-deformed_2021} as well as the Green-Schwarz sigma model \cite{costello_chern-simons_2020} on $\ads$ were also constructed using this framework. Nevertheless, none of the constructions included the ghost sector of the worldsheet action, which plays an essential role in the pure spinor formalism. 

The purpose of this work is to show how the 4DCS framework with gauge supergroup $\mathrm{PSU}(2,2|4)$ can be used to describe the complete $\ads$ pure spinor superstring including the ghost sector. In the construction of \cite{costello_gauge_2019}, there is some arbitrariness in the choice of pole structure for the Lax connection. In particular, one imposes by hand which components of $\Lax$ have poles at given points, giving rise to the chiral and anti-chiral defects. Here, we propose that this arbitrary choice should be interpreted as a gauge-fixing condition. We then use a set of $\mathrm{PSU}(2,2|4)$-valued meromorphic gauge transformations to restrict the pole structure of $\Lax$. 
 
Fixing the meromorphic gauge symmetry will lead to a set of 32 bosonic and 10 fermionic ghosts localized in two-dimensional subspaces of $\mathds{R}^2\times\mathds{CP}^1$. This characterizes the ghost system as a so-called ``order defect", and defects of this type were thoroughly analyzed in \cite{costello_gauge_2019}. Adapting a cohomology argument of \cite{aisaka_new_2003} to $\ads$, we argue that these ghosts should be equivalent to the usual pair of left and right-moving pure spinor bosonic ghosts, and the system composed by the 4DCS gauge field coupled to the ghosts yields the complete pure spinor $\ads$ superstring.
Thus, 4DCS not only gives a complete description of the formalism but also furnishes a natural origin for the pure spinor ghosts in terms of gauge-fixing. Note that in \cite{berkovits_simplifying_2009}, the 22 bosonic pure spinor ghosts in an $\ads$ background were similarly derived from 32 bosonic and 10 fermionic ghosts by gauge-fixing $\mathrm{PSU}(2,2|4)$ local symmetries.

The paper is organized as follows:  In the following subsection, we review the pure spinor formalism in $\ads$. In section \ref{CS_Sec} we introduce 4D Chern-Simons and the defects that lead to coset sigma models. In section \ref{gaugeFixSec} we discuss gauge invariance and argue that for a specific gauge-fixing, the Faddeev-Popov procedure leads to a ghost system and BRST charge which are equivalent to those of the pure spinor formalism. And in section \ref{sigmaModelAction}, we show how the 4D setup of the defects and ghosts leads to the 2D action and Lax connection of the pure spinor superstring in $\ads$. 
\subsection{Review of the pure spinor formalism in $\ads$}
We are interested in describing superstrings propagating in $\ads$. The associated superspace can be described in terms of the supercoset
\begin{equation}
    \frac{\mathrm{PSU}(2,2|4)}{\mathrm{SO}(4,1)\times\mathrm{SO}(5)}.\label{adsCoset}
\end{equation}

The $\mathfrak{psu}(2,2|4)$ algebra is given by
\begin{align}
& \left\{Q_\alpha, Q_\beta\right\}=\gamma_{\alpha \beta}^a P_{a},\quad\left\{\hat Q_{\hat{\alpha}}, \hat Q_{\hat{\beta}}\right\}=\gamma^a_{\hat{\alpha} \hat{\beta}} P_{a}, \label{QQcomm}\\
& {\left[P_{a}, \hat Q_{\hat{\beta}}\right]=-\frac{1}{2}\left(\eta \gamma_{a}\right)^\alpha_{\hat{\beta}} Q_\alpha,} \\
&{\left[P_{a}, Q_\alpha\right]=\frac{1}{2}\left(\gamma_{a} \eta\right)_\alpha^{\hat{\beta}} \hat{Q}_{\hat{\beta}},} \\
& \left[P_a, P_b\right]=-L_{a b}, 
\\
& \left\{Q_\alpha, \hat{Q}_{\hat{\beta}}\right\}=\frac{1}{2}\left(\gamma^{a b} \eta\right)_{\alpha \hat{\beta}} L_{a b} .
\end{align}
and the non-vanishing supertraces are
\begin{align}
& \str\left(P_{a} P_{b}\right)=\eta_{a b}, \quad \str\left(Q_\alpha \hat{Q}_{\hat{\beta}}\right)=-2 \eta_{\alpha \hat{\beta}}, \\
& \str\left(L_{a b} L_{c d}\right)=\eta_{a[b} \eta_{c] d .}.
\end{align}
 where $a=0,\ldots,9$ are tangent space vector indices and $\alpha/\hat\alpha=1,\ldots,16$ are ten-dimensional spinor indices.

This supercoset describes a ``semi-symmetric space" because the superalgebra $\mathfrak{psu}(2,2|4)$  has an automorphism $\rho$ which induces a $\mathds{Z}_4$ grading
\begin{equation}
    \mathfrak{g}=\mathfrak{g}^{(0)}\oplus\mathfrak{g}^{(1)}\oplus\mathfrak{g}^{(2)}\oplus\mathfrak{g}^{(3)}
\end{equation}
such that 
\begin{equation}
    \rho\cdot\mathfrak{g}^{(k)}=i^k\mathfrak{g}^{(k)}
\end{equation}
where \begin{align}
    &\mathfrak{g}^{(0)}=\mathrm{span}(L_{ab})\cong\mathfrak{so}(4,1)\oplus\mathfrak{so}(5)\\
    &\mathfrak{g}^{(1)}=\mathrm{span}(Q_\alpha)\\&\mathfrak{g}^{(2)}=\mathrm{span}(P_a)\\&\mathfrak{g}^{(3)}=\mathrm{span}(\hat{Q}_{\hat\alpha}).
\end{align}

Therefore, the superstring in this space can be described by the supergroup-valued degree of freedom
\begin{equation}
    g(w,\bar w)\in \mathrm{PSU}(2,2|4)
\end{equation}
with the equivalence relation 
\begin{equation}
    g(w,\bar w) \sim g(w,\bar w)h(w,\bar w) \;\;\;\;\; h \in \mathrm{SO}(4,1)\times\mathrm{SO}(5) \label{QuotientEquivalence}
\end{equation}
 where $(w, \bar w)$ are the complex coordinates of the worldsheet.
The matter sector of the action is constructed out of the $\mathfrak{psu}(2,2|4)$-valued left-invariant currents
\begin{equation}
    J:=g^{-1}d g
\end{equation}
which can be decomposed in terms of the $\mathds{Z}_4$ grading of $\mathfrak{psu}(2,2|4)$ as
\begin{equation}
    J=J^{(0)}+J^{(1)}+J^{(2)}+J^{(3)}.
\end{equation}
In particular, the matter part is \cite{mazzucato_superstrings_2012}
\begin{equation}
    S_{m}=\int d^2 w \;\str\big(\frac{1}{2}J_{\bar w}^{(2)}J_w^{(2)}+\frac{3}{4}J_w^{(3)} J_{\bar w}^{(1)}+\frac{1}{4}J_w^{(1)}J_{\bar w}^{(3)}\big). \label{PS_Matter}
\end{equation}
We also have ghosts which play a fundamental role in cancelling the conformal anomaly and constructing physical vertex operators. As in 10D flat space, they are bosonic fields with 10D Weyl spinor indices ($\lambda^\alpha$,\;$\hat\lambda^{\hat\alpha}$) that satisfy the pure spinor constraints
\begin{align}
    \lambda^\alpha\gamma^a_{\alpha\beta}\lambda^\beta=0\\
    \hat\lambda^{\hat \alpha}\gamma^a_{\hat \alpha \hat \beta}\hat\lambda^{\hat\beta}=0.
\end{align}
In the $\ads$ model, they can be conveniently written in terms of $\mathfrak{g}$-valued ghosts as
\begin{align}
    &\lambda:= \lambda^\alpha Q_\alpha \in \mathfrak{g}^{(1)}\\
    &\hat\lambda:=\hat\lambda^{\hat \alpha}\hat{Q}_{\hat\alpha} \in \mathfrak{g}^{(3)}.
\end{align}
From the structure constants in (\ref{QQcomm}), it is direct to see that the pure spinor constraints can be rewritten as
\begin{equation}
    \{\lambda,\lambda\}=\{\hat\lambda,\hat\lambda\}=0. \label{algebraicPS}
\end{equation}
The model also has anti-ghosts $(y,\hat{y})\in (\mathfrak{g}^{(3)},\mathfrak{g}^{(1)})$ which can be used to construct the ghost-number zero objects
\begin{equation}
    N=-\{y,\lambda\}\;\;\;\;\;\hat{N}=-\{\hat{y},\hat\lambda\}\;\;\in\mathfrak{g}^{(0)} \label{ghostCurrents}
\end{equation}
The ghost contribution to the action is then
\begin{equation}
    S_{gh}=\int d^2 w \;\str\big(y\nabla_{\bar w}^{(0)}\lambda +\hat{y}\nabla^{(0)}_w\hat\lambda-N\hat{N}\big),
\end{equation}
where we defined the $\mathfrak{g}^{(0)}$-covariant derivative 
\begin{equation}
\nabla^{(0)}=d+[J^{(0)},\cdot].
\end{equation}
The complete action is just the sum
\begin{equation}
    S=S_{m}+S_{gh.}.
\end{equation}
This action is invariant under the BRST transformations
\begin{align}
    &\;\;\;\;\;\;\;Q\cdot g=g(\lambda+\hat\lambda)\\
    &Q\cdot y=-J^{(3)}_w\;\;\;\;Q\cdot\hat{y}=-J^{(1)}_{\bar w}\\
    &\;\;\;\;\;\;Q\cdot\lambda=Q\cdot\hat{\lambda}=0
\end{align}
and the Noether procedure yields the BRST charge 
\begin{align}
    Q=\oint dw\;\str\big(\lambda J^{(3)}_{\bar w}\big)-\oint d\bar w\;\str\big(\hat\lambda J^{(1)}_w\big). \label{PureSpinorBRST}
\end{align}

In this sigma model, the equations of motion are equivalent to the flatness of the Lax connection
\begin{align}
    &\Lax_w=J_w^{(0)}+\frac{1}{u}J_w^{(1)}+\frac{1}{u^2}J_w^{(2)}+\frac{1}{u^3}J_w^{(3)}-(1-\frac{1}{u^4})N\\
    &\Lax_{\bar w}=J_{\bar w}^{(0)}+u J_{\bar w}^{(3)}+u^2 J_{\bar w}^{(2)}+u^3 J_{\bar w}^{(1)}-(1-u^4)\hat{N}.
\end{align}
where $u$ is the spectral parameter defined in (\ref{LaxFlat}).
Moreover, one can show \cite{mikhailov_perturbative_2007} that, under BRST transformations, the Lax connection transforms as
\begin{equation}
    Q\cdot \Lax =  d\tilde \lambda +[\Lax, \tilde \lambda] \label{DressingTransformation}
\end{equation}
where 
\begin{equation}
    \tilde \lambda =\frac{1}{u}\lambda+u\hat{\lambda}. \label{DressingParameter}
\end{equation}

Note that (\ref{DressingTransformation}) has exactly the form of the gauge transformation in Chern-Simons theory. Thus, it seems reasonable that the ghosts of the formalism can be reinterpreted as Faddeev-Popov ghosts in a gauge-fixing of 4D Chern-Simons theory. Moreover, from the form of (\ref{DressingParameter}) it is natural to expect that they are related to meromorphic gauge transformations. This will be shown below by starting with the 4DCS action of \cite{costello_gauge_2019} for the $\ads$ superstring and gauge-fixing the local symmetries.
\section{4D Chern-Simons} \label{CS_Sec}
Our starting point is the 4D Chern-Simons (4DCS) action given by \cite{costello_gauge_2018,costello_gauge_2018-1,costello_gauge_2019}
\begin{equation}
    S=\frac{1}{2\pi i}\int_\mathcal{M}\omega \wedge \str\big(A\wedge dA+\frac{1}{3}A\wedge\big[A\wedge A\big]\big)
    \label{CSaction}
\end{equation}
with a gauge (super)group $G=\mathrm{PSU}(2,2|4)$ and $\mathcal{M}=\Sigma\times\mathds{CP}^1$, where $\Sigma$ is the two-dimensional worldsheet and we set $\Sigma=\mathds{R}^2$ from now on. Throughout the paper, ($z, \bar z$) are coordinates on $\mathds{CP}^1$ and $(w,\bar w)$ are coordinates on $\mathds{R}^2$.
Moreover, $\omega(z)$ is a meromorphic 1-form on $\mathds{CP}^1$
\begin{equation}
    \omega:=\phi(z)dz.
\end{equation}
which can have poles and zeroes at special points. These special points are called disorder defects \cite{costello_gauge_2019} and play an important role in the relation between 4DCS and 2D integrable field theories. One can construct a plethora of known integrable models by choosing different functions $\phi(z)$ and gauge groups $G$.

The equation of motion for $A$ is
\begin{align}
    \omega\wedge F(A)=0 \label{EOM}
\end{align}
where the field-strength $F$ is defined as 
\begin{equation}
    F(A):= dA+A\wedge A.
\end{equation}
\subsection{Coset models and $\ads$}

\begin{figure}
    \centering
    \begin{tikzpicture}[dot/.style={circle,inner sep=1pt,fill,label={#1},name=#1},
        extended line/.style={shorten >=-#1,shorten <=-#1},
        extended line/.default=1cm]
        \draw (0,0) rectangle (5,5);
        \draw (1, 2.5) -- (4, 2.5);
        \node [dot={$z=0$}] at (1,2.5){};
        \node [dot={$z=\frac{1}{4}$}] at (4,2.5){};
        \node [draw] at (4.25,.3) {{$z$-plane}};
        \draw [->] (2.5,2) to [out=10,in=-10] (2.5,3);
        \node [] at (2.3,1.9) {{$A$}};
        \node [] at (2.1,3.2) {{$\rho(A)$}};
        \draw[->] (2.5,-.5) -- (2.5,-1.5);
        \begin{scope}[shift={(-8,-7)}]
        \draw (8,0) rectangle (13,5);
        \draw (8,0) -- (13,5);
        \draw (8,5) -- (13,0);
        \node [anchor=west] at (10.6,2.5) {{$u=0$}};
        \node [dot={}] at (10.5,2.5) {};
        \filldraw[color=white] (11.5,0) rectangle (13,.6);
        \node [draw] at (12.25,.3) {{$u$-plane}};
        \draw [->] (12.15, 2.15) to [out=-90,in=0] (11, 1.25);
        \draw [->] (10, 1.25) to [out=180,in=-90] (8.85, 2.15);
        \draw [->] (8.85, 2.85) to [out=90,in=180] (10, 3.75);
        \draw [->] (11, 3.75) to [out=0,in=90] (12.15, 2.85);
        \node [] at (12. 15,2.5) {{$A$}} ;
        \node [] at (10.5, 1.25) {{$\rho(A)$}}; 
        \node [] at (8.85, 2.5) {{$\rho^2(A)$}}; 
        \node [] at (10.5, 3.75) {{$\rho^3(A)$}}; 
        \end{scope}
    \end{tikzpicture}
    \caption{$z$-plane with line defect and change of coordinates to 4-fold cover. The line defect correspond to the 4 lines connecting $u=0$ and $u=\infty$ in the second diagram.}
    \label{fig:DefectDiag}
\end{figure}
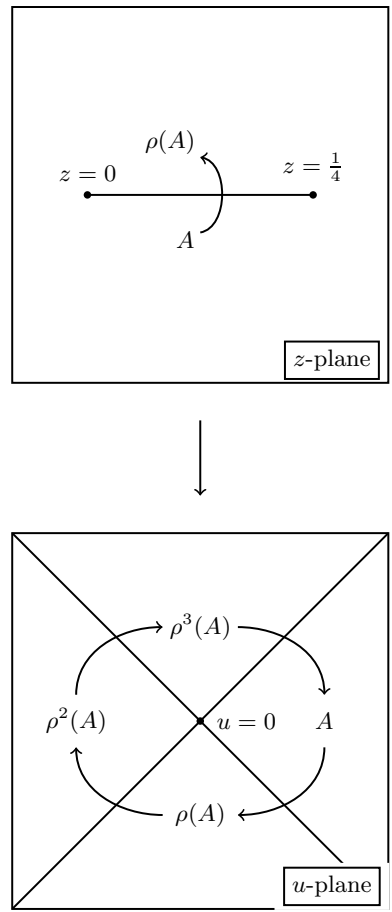

To describe sigma-models with target space $\ads$, we need to introduce another kind of defect \cite{costello_gauge_2019} which is a line with ends at $z=0$ and $z=\frac{1}{4}$ such that, when the line defect is crossed, the $\mathds{Z}_4$ automorphism $\rho$ of the superalgebra is applied on the gauge field, as shown in fig. \ref{fig:DefectDiag}.

It is easier to understand the implications of this defect in a 4-fold cover of $\mathbb{CP}^1$ defined by the change of coordinates.
\begin{equation}
     z=-\frac{1}{4(u^4-1)}.
\end{equation}
The holomorphic 1-form $dz$ is then 
\begin{equation}
    dz=\frac{u^{3}}{(u^4-1)^2}du=:\omega(u) .\label{omega}
\end{equation}
The 1-form has zeroes at $u=0$ and $u=\infty$ (the pre-images of the ends of the line defect) and second-order poles at $\{p_k\}=\{1,i,-1,-i\}$. In addition to understanding the behavior of the theory at these poles and zeroes, we also need to deal with the line defect. In particular, we shall impose that the fields are single-valued on the $z$-plane. In the $u$-plane, this implies that field configurations should be invariant under the simultaneous permutation of the 4 ``slices" of the cover and application of $\rho$, as can be seen in the second diagram of fig. \ref{fig:DefectDiag}. This is true if, and only if,
\begin{align}
    A(u)=\rho\cdot A(iu), \label{RhoConstraintA}
\end{align}
and we shall call (\ref{RhoConstraintA}) the $\rho$-constraint.

Such a configuration of defects was used to construct the matter sector of the $\ads$ pure spinor superstring in \cite{costello_gauge_2019}. 
As we will see in the following sections, the equations of motion imply that the gauge field along $\Sigma$ can have poles either at $u=0$ or $u=\infty$. In the construction of \cite{costello_gauge_2019}, one imposes by hand that $A_w$ only has poles at $u=0$ and that $A_{\bar{w}}$ only has poles at $u=\infty$. However, there is no \textit{a priori} reason for the Lax connection to satisfy this property, and we will argue below that this should be interpreted as a gauge-fixing condition. We will see in section \ref{gaugeFixSec} that after appropriately imposing this gauge-fixing, the usual left- and right-moving pure spinor ghosts emerge as Faddeev-Popov ghosts.
\subsection{Boundary conditions} \label{BCsection}
In this theory, it is convenient to consider small contours around the poles $p_k=\{1,i,-1,-i\}$ of $\omega(u)$ as boundaries. We shall impose Dirichlet boundary conditions on the gauge field $A_\Sigma (p_k)$ at these contours so that the boundary terms in the variation of the action vanish. Note that by varying the action, we get
\begin{align}
    \delta S=\frac{1}{2\pi i}\sum_k\int_{\Sigma}\oint_{\mathcal{C}_k}\omega(u)\wedge\str\big(\delta A_\Sigma\wedge A_\Sigma\big)+\nonumber
    \\
    +\frac{1}{\pi i}\int_{\Sigma\times\mathds{CP}^1}\omega \wedge\str\big(\delta A\wedge F(A)\big) \label{GenericVariation}
\end{align}
where the index $k$ goes over the boundaries. So the boundary variation vanishes if we impose the Dirichlet boundary condition
\begin{equation}
    A_\Sigma(p_k)=0. \label{Dirichlet}
\end{equation}
\section{Gauge invariance and gauge fixing}\label{gaugeFixSec}
The action is invariant under the gauge symmetry 
\begin{equation}
    A\mapsto h A h^{-1}-dh h^{-1} \;\;\;\; h\in\mathrm{PSU}(2,2|4) \label{gaugeS}
\end{equation}
provided that condition (\ref{Dirichlet}) is preserved at $u^4=1$. This means that the $\bar u$ component of $A$ can be gauged to zero everywhere except near the points $u^4 =1$, which leads to $G$-valued degrees of freedom at the vicinity of $u^4=1$
    \begin{align}
        &A_{\bar u}=-\partial_{\bar u}\hat{g}\hat{g}^{-1}\label{hatG}\\
        &A_{\Sigma}=\hat{g}\mathcal{L}\hat{g}^{-1}-d_{\Sigma}\hat g\hat{g}^{-1}\label{L}
    \end{align}
where $\mathcal{L}$ is a 1-form with components only along $\Sigma\equiv\mathds{R}^2$.

The Dirichlet boundary conditions on $A_\Sigma$ translate to
\begin{align}
    \Lax(p_k)=\hat{g}^{-1}d\hat g|_{u=p_k}.
    \label{LaxBC}
\end{align}
In terms of ($\Lax,\hat{g}$) the 4DCS action (\ref{CSaction}) is 
\begin{align} 
    S(\Lax,\hat{g})=&\frac{1}{2\pi i}\int_{\mathcal{M}}\omega\wedge\str\big(\Lax\wedge \bar\partial\Lax\big)+
    \nonumber\\
    &+\frac{1}{2\pi i}\sum_k \oint_{\mathcal{C}_k}\omega(u)\int_\Sigma \str\big(\Lax\wedge\hat{g}^{-1}d\hat{g}\big)+\nonumber\\
    &+\frac{1}{6\pi i}\int_{\mathcal{M}}\omega\wedge\str\big(\hat{g}^{-1}d\hat{g}\wedge\hat{g}^{-1}d\hat{g}\wedge\hat{g}^{-1}d\hat{g}\big). \label{gHatLAction}
\end{align}

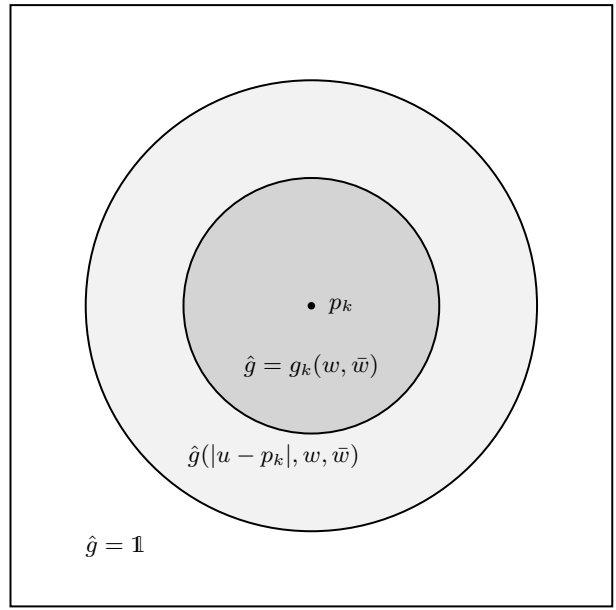
\begin{figure}
    \centering
    \begin{tikzpicture}[dot/.style={circle,inner sep=1pt,fill,label={#1},name=#1}]
        \draw (0,0) rectangle (8,8);
        \filldraw[fill=black!5] (4,4) circle (3);
        \filldraw[fill=black!17] (4,4) circle (1.7);
        \node [dot] at (4,4) {};
        \node [] at (4,3.2) {{$\hat{g}=g_k(w,\bar w)$}};
        \node [] at (4.4, 4) {{$p_k$}};
        \node [] at (3.5, 2) {{$\hat{g}(|u-p_k|,w,\bar w)$}};
        \node [] at (1.4,.8) {{$\hat{g}=\mathds{1}$}};
    \end{tikzpicture}
    \caption{Schematic depiction of archipelago gauge conditions.}
    \label{fig:archipelagoFigure} 
\end{figure}

To get an $A_{\bar u}$ that satisfies the desired gauge-fixing conditions, we use (\ref{gaugeS}) to fix $\hat{g}$ to be \textit{archipelago-like}\cite{delduc_unifying_2020}. It will then be equal to $\mathds{1}$ outside small disks around the poles $\{p_k\}$. Inside the disks, there is a smaller disk in which $\hat{g}=g_k(w,\bar w)$ (for the disk around $p_k$) and an outer annulus in which $\hat{g}$ smoothly interpolates between $g_k$ and $\mathds{1}$ depending only on $|u-p_k|$ and $(w,\bar w)$. These conditions are schematically depicted in Fig. \ref{fig:archipelagoFigure}. Note that the $\rho$-constraint fixes $g_k=\rho^k \cdot g$ where we defined $g_0\equiv g$.
\subsection{Residual gauge transformations and ghosts}
In archipelago gauge, the equation of motion (\ref{EOM}) implies
\begin{align}
    &\phi(u)\partial_{\bar u}\Lax=0\label{Lpoles}\\
    &\phi(u)(d\Lax+\Lax\wedge\Lax)=0.
\end{align}
where comparing with (\ref{omega}) we see that
\begin{align}
    \phi(u):=\frac{u^3}{(u^4-1)^2}.
\end{align}

The residual gauge transformations should be consistent with these equations and leave $A_{\bar u}$ unchanged. Since (\ref{Lpoles}) implies that $\Lax$ only has poles up to third-order, the residual transformations are
\begin{align}
    \delta \Lax = \nabla_\Sigma \tilde\Lambda
\end{align}
where 
\begin{equation}
    \nabla_\Sigma :=d_\Sigma + [\Lax, \cdot],
\end{equation}
and $\tilde{\Lambda}$ should be defined such that $\delta\Lax$ only has poles up to third order at $u=0$ or $u=\infty$ and should vanish at $u^4=1$ to preserve the boundary condition (\ref{LaxBC}). The precise form of these parameters will depend on the specific value of $\Lax$, but the general structure is 
\begin{widetext}
\begin{align}
    \tilde\Lambda = \big(1/u-u^3\big)\Lambda_1(w,\bar w) +\big(1/u^2-u^2\big)\Lambda_2(w,\bar w)+\big(u-1/u^3\big)\Lambda_3(w,\bar w) + \mathrm{higher}\;\mathrm{poles} \label{tildeLambda}
\end{align}
\end{widetext}
where $\Lambda_k\in\mathfrak{g}^{(k)}$. Here, ($\Lambda_1$, $\Lambda_2$, $\Lambda_3$) are independent parameters and the higher poles are determined in terms of ($\Lambda_1$, $\Lambda_2$, $\Lambda_3$) by the requirement that $\nabla_\Sigma \tilde\Lambda$ only has poles up to third order. We therefore have 32 fermionic ($\Lambda_1$ and $\Lambda_3$) and 10 bosonic ($\Lambda_2$) gauge parameters. Since each fermionic/bosonic gauge parameter leads to a bosonic/fermionic ghost, a naive counting indicates that the ghost system obtained from gauge-fixing this symmetry is equivalent to $32-10=22$ bosonic ghosts, where the fermionic ghosts have been interpreted as ghosts-for-ghosts which cancel 10 of the bosonic ghosts. This is precisely the number of degrees of freedom of a pair of pure spinors, and a similar derivation of 22 pure spinor bosonic ghosts from 32 bosonic and 10 fermionic ghosts was used in \cite{berkovits_simplifying_2009}.

To further understand this, let's use $(\Lambda_1, \Lambda_2, \Lambda_3)$ to impose gauge-fixing conditions on $\Lax$. Let
\begin{equation}
    \Lax=\sum_k u^k \Lax^k.
\end{equation}
We can first use $\Lambda_1$ and $\Lambda_3$ to impose 
\begin{equation}
    \Lax_w^{1}=\Lax_{\bar w}^{-1}=0,
    \label{FermionicGF}
\end{equation}
which leads to a pair of ghosts $(Z,\hat Z)\in(\mathfrak{g}^{(1)},\mathfrak{g}^{(3)})$. We can then use $\Lambda_2$ to gauge away 5 components of $\Lax_w^2$ and $\Lax_{\bar w}^{-2}$. To choose which components are gauged away, we follow the procedure of \cite{berkovits_simplifying_2009} by defining the matrices ($\mathcal{N}$, $\bar{\mathcal{N}}$) as in Appendix \ref{PsAppendix} and imposing 
\begin{equation}
    \mathcal{N}(Z)\cdot\Lax^{2}_w=\bar{\mathcal{N}}(\hat{Z})\cdot\Lax_{\bar w}^{-2}=0 .\label{BosonicGF}
\end{equation}
Given these conditions, one can derive further restrictions from the flatness equation. The vanishing of fourth- and fifth-order poles at $u=0$ of the curvature implies 
\begin{align}
    &[\Lax_{w}^{-1},\Lax_{\bar w}^{-3}]+[\Lax_{w}^{-2},\Lax_{\bar w}^{-2}]=0\\
    &[\Lax_{w}^{-2},\Lax_{\bar w}^{-3}]+[\Lax_{w}^{-3},\Lax_{\bar  w}^{-2}]=0
\end{align}
which in components gives
\begin{align}
    &L_{ab}[(\Lax^{-1}_w)^\alpha(\Lax_{\bar w}^{-3})^{\hat\beta}(\gamma^{ab}\eta)_{\alpha\hat{\beta}}+(\Lax_w^{-2})^a\eta^{bc}\mathcal{N}^I_c\bar\Lax_{I}]=0 \label{fourthPoleCurvature}\\
    &Q_\alpha[(\Lax^{-2}_w)^a(\eta\gamma_a)_{\hat\beta}^\alpha(\Lax_{\bar w}^{-3})^{\hat\beta}+\eta^{ab}\mathcal{N}^I_b\bar \Lax_{I}(\eta\gamma_a)^\alpha_{\hat\beta}(\Lax_w^{-3})^{\hat\beta}]=0\label{fifthPoleCurvature}
\end{align}
where we used condition (\ref{BosonicGF}) to write 
\begin{equation}
    (\Lax_{\bar w}^{-2})^a=\eta^{ab}\mathcal{N}^I_b\bar\Lax_I .
\end{equation}
Assuming that $\Lax_w^{-1}$ and half of $\Lax_w^{-2}$ are non-zero and generic (as restricting their values would correspond to other gauge-fixing conditions), equations (\ref{fourthPoleCurvature}) and (\ref{fifthPoleCurvature}) imply 
\begin{equation}
    \Lax_{\bar w}^{-3}=\bar\Lax_I=0.
\end{equation}
So we have shown that in this gauge, $\Lax_{\bar w}^{-1} = \Lax_{\bar w}^{-2} = \Lax_{\bar w}^{-3}=0$, i.e. $\Lax_{\bar w}$ has no poles at $u=0$.
An analogous argument considering the poles of $F(\Lax)$ at $u=\infty$ implies that $\Lax_w$ has no poles at $u=\infty$.

To summarize, we have argued that some of the conditions on the pole structure of $\Lax$ come from a gauge-fixing choice and the remaining conditions come from consistency with the flatness equation of motion. If one had instead tried to impose the conditions on the pole structure of $\Lax$ without a gauge-fixing choice (for example, by requiring that the action is finite near the zeros of $\omega$), one would not be able to uniquely fix the desired pole structure of $\Lax$.

From the form of the gauge transformation (\ref{tildeLambda}), the ghost system can be conveniently organized in the object
\begin{widetext}
\begin{align}
    \tilde C = \big(1/u-u^3\big)Z(w,\bar w) +\big(1/u^2-u^2\big)c(w,\bar w)+\big(1/u^3-u\big)\hat Z(w,\bar w) + \mathrm{higher}\;\mathrm{poles},
\end{align}
\end{widetext}
where $Z=Z^\alpha Q_\alpha $ and $\hat{Z}=Z^{\hat\alpha}\hat Q_{\hat{\alpha}}$ are the 32 bosonic ghosts, $c=c^aP_a$ are the 10 fermionic ones, and the higher poles are complicated functions of ($Z$, $c$, $\hat{Z}$).
Analogously, we have 32 bosonic anti-ghosts ($y_\alpha$, $\hat{y}_{\hat{\alpha}}$) associated to the gauge-fixing conditions (\ref{FermionicGF}) and 10 fermionic ones ($b_I$, $\bar b_I$) associated to (\ref{BosonicGF}). We also need the Lagrange multipliers ($l_\alpha$, $\hat {l}_{\hat{\alpha}}$, $f_I$, $\bar{f}_I$) to impose the gauge-fixing conditions. All these new fields can be conveniently organized in the objects
\begin{align}
    &B=y+u b\cdot\bar{\mathcal{N}}\\
    &\bar B=\hat{y}+\frac{1}{u}\hat b\cdot{\mathcal{N}}\\
    &\mathcal{F}=l+uf\cdot\bar{\mathcal{N}}\\
    &\bar{\mathcal{F}}=\hat{l}+\frac{1}{u}\bar{f}\cdot\mathcal{N}.
\end{align}
The Faddeev-Popov action can then succinctly be written as
\begin{align}
    S_{FP}=\int_\Sigma d^2 w\bigg\{&\oint_{\mathcal{C}_0}du\;\str\big(B\nabla_{\bar w}\tilde C+\mathcal{F}\Lax_{\bar w}\big)+\nonumber\\
    +&\oint_{\mathcal{C}_{\infty}}du\;\str\big(\bar B\nabla_w\tilde{C}+\bar{\mathcal{F}}\Lax_w\big)\bigg\} \label{ComponentGhostAction1}
\end{align}
The BRST currents can be easily derived through the Noether procedure, and the left-moving component is
\begin{align}
    j_w^{B}=&\oint_{\mathcal{C}_0} du \;\str\big(\mathcal{F}\tilde C-\frac{1}{2}B\{\tilde C,\tilde C\}\big)\nonumber\\
    =&l_\alpha Z^\alpha+f_I\mathcal{N}^I_ac^a -\frac{1}{2}y_\alpha\{\tilde C,\tilde C\}_{-1}^\alpha-\frac{1}{2}b_I\bar{\mathcal{N}}^I_a\{\tilde{C},\tilde{C}\}^a_{-2}
\end{align}
where $\{\tilde C,\tilde C\}_k$ denotes the $k$-th Laurent coefficient of $\{\tilde C, \tilde C\}$.
Note that 
\begin{align}
    &\{\tilde C,\tilde C\}_{-1} = -\{c,\hat Z\}+\;\mathrm{higher}\;\mathrm{pole}\;\mathrm{contrib.}\\
    &\{\tilde C, \tilde C\}_{-2} = -\frac{1}{2}\{Z,Z\}-\frac{1}{2}\{\hat{Z},\hat{Z}\}+\;\mathrm{higher}\;\mathrm{pole}\;\mathrm{contrib.}
\end{align}
The left-moving BRST current is then
\begin{align}
    j_w^B=&l_\alpha Z^\alpha+f_I\mathcal{N}^I_ac^a + \frac{1}{4}y_\alpha (\eta\gamma_a)^\alpha_{\hat\beta}c^a\hat Z^{\hat \beta}+\nonumber\\
    &-\frac{1}{2}b_I\bar{\mathcal{N}}^I_a\eta^{ab}\big(\mathcal{N}^J_b\Phi_J(Z)+\bar{\mathcal{N}}^J_b\Phi_J(\hat Z)\big)+\ldots 
\end{align}
where we omitted the higher pole contributions.
Then using (\ref{nullIdentity}) for $\bar {\mathcal{N}}$ yields
\begin{align}
    j_w^B=&l_\alpha Z^\alpha+f_I\mathcal{N}^I_ac^a + \frac{1}{4}y_\alpha (\eta\gamma_a)^\alpha_{\hat\beta}c^a\hat Z^{\hat \beta}+\nonumber\\
    &-\frac{1}{2}b_I\bar{\mathcal{N}}^I_a\eta^{ab}\mathcal{N}^J_b\Phi_J(Z)+\;\mathrm{higher}\;\mathrm{pole}\;\mathrm{contributions}.\label{preB}
\end{align}
This is the BRST current of \cite{berkovits_simplifying_2009} plus contributions from higher-order poles in $\tilde{C}$. 

Following \cite{berkovits_simplifying_2009}, this BRST charge is an $\ads$ generalization of the BRST charge in flat space of Aisaka and Kazama \cite{aisaka_new_2003}. In \cite{aisaka_new_2003}, Aisaka and Kazama used homological perturbation theory to show that the cohomology of the BRST charge 
\begin{align}
    Q = \oint dw (\;d_\alpha Z^\alpha + P_I \mathcal{N}_a^I c^a + ...) \label{aisaka2}
\end{align} 
where $Z^\alpha$ is unconstrained and $(d_\alpha, P_I)$ are fermionic and bosonic operators in flat space, is equivalent to the cohomology of the BRST charge
\begin{align}
        Q_{\lambda,\hat\lambda}=\oint dw \;d_\alpha \lambda^\alpha 
\end{align}
where $\lambda^\alpha$ is a pure spinor.
Using similar arguments to those of \cite{aisaka_new_2003}, it should be possible to show that the cohomology of $Q = \oint dw \;j_w^B +\oint d\bar w\; \hat j_W^B$ where $j_W^B$ is defined in (\ref{preB})
is equivalent to the cohomology of the charge
\begin{align}
        Q_{\lambda,\hat\lambda}=\oint dw \;l_\alpha \lambda^\alpha +\oint d\bar w \;\hat{l}_{\hat \alpha}\hat\lambda^{\hat\alpha}, \label{psBRST}
\end{align}
where the ghost system is now composed of a pair of pure spinors ($\lambda$,$\hat\lambda$). Although the terms $...$ in (\ref{aisaka2}) are simpler than the higher pole contributions in (\ref{preB}), it is expected that homological perturbation theory can similarly be used to argue that the higher pole contributions do not affect the cohomology of the BRST charge.

The associated Faddeev-Popov action related to (\ref{psBRST}) is 
\begin{align}
    S_{\lambda, \hat\lambda}= -2\bigg(&\oint_{\mathcal{C}_0} du \int_\Sigma d^2w \;Q\cdot\str\big(B\Lax_{\bar w}\big)+\nonumber\\
    +&\oint_{\mathcal{C}_\infty} d u \int_\Sigma d^2w \;Q\cdot\str\big(\bar B\Lax_{w}\big)\bigg),
\end{align}
where the overall normalization was chosen for later convenience.
With this new BRST charge, the BRST transformation of the Lax connection is 
\begin{equation}
    Q\cdot\Lax =  \nabla_\Sigma \big(\frac{1}{u}\lambda + u \hat\lambda\big),
\end{equation}
which can be verified \textit{a posteriori} through the Noether procedure. Thus, the action for the pure spinor ghosts is
\begin{equation}
    S_{\lambda,\hat \lambda} = 2\int_\Sigma d^2w\;\str\big(y\nabla^0_{\bar w}\lambda+\hat{y}\nabla_w^0\hat \lambda - l\Lax_{\bar w}^{-1}-\hat l \Lax_w^1\big),
\end{equation}
where we already used the multipliers ($f_I$,$\bar f_I$) to impose the conditions (\ref{BosonicGF}). Moreover, note that since the BRST transformation of $\Lax$ no longer vanishes at $u^4=1$, we also need to define $Q\cdot g=g(\lambda+\hat\lambda)$ to respect the boundary conditions of (\ref{Dirichlet}).
\section{The 2D sigma model from 4D action}\label{sigmaModelAction}
The gauge-fixed action is then obtained by the sum of (\ref{gHatLAction}) with the Faddeev-Popov ghosts and the contributions from the Lagrangian multipliers 
\begin{align}
    S=&\frac{1}{2\pi i}\int_{\mathcal{M}}\omega\wedge\str\big(\Lax\wedge\bar \partial \Lax\big)+\nonumber\\
    &+\frac{1}{2\pi i}\sum_k\oint_{\mathcal{C}_k}\omega(u)\int_\Sigma\str\big(\mathcal{L}\wedge \hat g^{-1}d\hat g\big)+\nonumber\\
    &+\frac{1}{6\pi i}\int_{\mathcal{M}}\omega\wedge\str\big(\hat{g}^{-1}d\hat{g}\wedge\hat{g}^{-1}d\hat{g}\wedge\hat{g}^{-1}d\hat{g}\big)+\nonumber\\
    &+2\int_\Sigma d^2w\;\str\big(y\nabla^{0}_{\bar w}\lambda+\hat{y}\nabla_{w}^{0}\hat \lambda-l\Lax_{\bar w}^{-1}  -\hat l\Lax^{1}_w \big). \label{CS+Ghosts}
\end{align}

We shall now vary the action with respect to $\Lax$ and use the resulting equations of motion, gauge fixing conditions and boundary conditions $\Lax(p_k)=\rho^k\cdot J$ to solve for $\Lax$ in terms of $J$ and the ghosts. We then plug $\Lax$ back in (\ref{CS+Ghosts}) to obtain the action for the associated 2D integrable field theory. As we will see, the resulting action and Lax connection describe the pure spinor superstring in $\ads$. The resulting model is BRST invariant and we show that the BRST charge (\ref{psBRST}) coincides with the usual one in the pure spinor formalism by writing the Lagrange multipliers ($l_\alpha, \hat l _{\hat \alpha}$) in terms of physical fields.
\subsection{Equations of motion and Lax connection} 
First of all, note that the WZ-term in the second line of (\ref{CS+Ghosts}) vanishes for the 1-form (\ref{omega}) since for archipelago-like $\hat g$
\begin{align}
    &\int_{\mathcal{M}}\omega\wedge\str\big(\hat{g}^{-1}d\hat{g}\wedge\hat{g}^{-1}d\hat{g}\wedge\hat{g}^{-1}d\hat{g}\big)\propto\nonumber\\
    &\sum_k\res{(\phi,p_k)}\int_{\Sigma\times\mathcal{A}_k}\str\big(J\wedge J\wedge J\big) \label{WouldBeWZ}
\end{align}
where $\mathcal{A}_k$ denotes the small annulus around $p_k$ in which $\hat g$ depends on $|u-p_k|$. As $\phi(u)$ only has double poles, all terms on the right-hand side of (\ref{WouldBeWZ}) are zero.
Moreover, the second term in (\ref{CS+Ghosts}) becomes
\begin{align}
    &\frac{1}{2\pi i}\sum_k\oint_{\mathcal{C}_k}\omega(u)\int_\Sigma\str\big(\mathcal{L}\wedge \hat g^{-1}d\hat g\big)=\nonumber\\&\frac{1}{2\pi i}\sum_k\oint_{\mathcal{C}_k}\omega(u)\int_\Sigma\str\big(\mathcal{L}\wedge \rho^k\cdot(g^{-1}d g)\big).
\end{align}
A special feature of (\ref{CS+Ghosts}) is that the ghosts only couple to $\Lax$ at the points $u=\{0,\infty\}$, which means that they ``source" higher poles for the Lax connection at these points. To see this, let's vary the action with respect to $\Lax$
\begin{widetext}
\begin{align}
    \delta S=&\frac{1}{2\pi i}\int_{\mathcal{M}}\omega\wedge\str\big\{\delta\Lax\wedge\bar \partial\Lax+\Lax\wedge\bar \partial(\delta \Lax)\big\}+2\int_{\mathcal{M}}d^2wd^2u\;\str\big(\delta_{u,0} N \delta\Lax_{\bar w}^{(0)}+\delta_{u,\infty}\hat{N}\delta\Lax^{(0)}_{w}\big)\nonumber\\
    =&\frac{1}{\pi i}\int_{\mathcal{M}}\omega\wedge\str\big(\delta\Lax\wedge\bar d\Lax\big)+\frac{1}{2\pi i}\sum_k\oint_{\mathcal{C}_k}\omega(u)\int_\Sigma\str\big(\Lax\wedge\delta\Lax\big)+\nonumber\\
    &+2\int_{\mathcal{M}}d^2wd^2u\;\str\big(\delta_{u,0} N \delta\Lax_{\bar w}^{(0)}+\delta_{u,\infty}\hat{N}\delta\Lax^{(0)}_{w}\big)\nonumber\\
    =&\frac{1}{\pi i}\int_{\mathcal{M}}\omega\wedge\str\big(\delta\Lax\wedge\bar d\Lax\big)+2\int_{\mathcal{M}}d^2wd^2u\;\str\big(\delta_{u,0} N \delta\Lax_{\bar w}^{(0)}+\delta_{u,\infty}\hat{N}\delta\Lax^{(0)}_{w}\big).
\end{align}
\end{widetext}
The boundary terms cancel due to the boundary condition $\Lax(p_k)=\rho^k\cdot J$ and $N$ and $\hat N$ are defined in (\ref{ghostCurrents}).
Therefore, the equations of motion for $\Lax$ are 
\begin{align}
    &\frac{1}{2\pi i}\phi(u)\partial_{\bar u}\Lax_w=-N\delta_{u,0} \label{LEOM} \\
        &\frac{1}{2\pi i}\phi(\xi)\partial_{\bar \xi}\Lax_{\bar w}=-\hat{N}\delta_{\xi, 0}\label{LbarEOM}
\end{align}
where we used the coordinates $\xi=\frac{1}{u}$ centered at infinity\footnote{The delta function $\delta_{u,\infty}$ is defined such that $d^2 u \delta_{u,\infty}=d^2\xi\delta_{\xi,0}$.} to write the e.o.m. for $\Lax_{\bar w}$.
Using the identity
\begin{equation}
    \partial_{\bar u}\big(\frac{1}{u^n}\big)=(-1)^n\frac{2\pi i}{(n-1)!}\partial_u^{(n-1)}\delta_{u,0},
    \label{deltaId}
\end{equation}
the e.o.m. imply that the fourth-order poles in $\Lax$ have the coefficients
\begin{align}
    \mathcal{L}_w^{-4}=N\\
    \mathcal{L}_{\bar w}^{4}=\hat{N}
\end{align}
Now, joining the e.o.m., the boundary conditions, the gauge fixing conditions and the $\rho$-constraint, $\Lax$ is 
\begin{align}
    &\mathcal{L}_w=J_w^{(0)}+u^{-1}J_{w}^{(1)}+u^{-2}J_{ w}^{(2)}+u^{-3}J_w^{(3)}+(u^{-4}-1)N \label{psLax}\\
    &\Lax_{\bar w}=J_{\bar w}^{(0)}+u J_{\bar w}^{(3)}+u^2 J_{\bar w}^{(2)}+u^3 J_{\bar w}^{(1)}+(u^4-1)\hat{N}, \label{psBarLax}
\end{align}
This is, indeed, the Lax connection for the pure spinor formalism in $\ads$ \cite{magro_review_2012}. Note that after gauge-fixing, this is the unique solution with poles up to third order in the Lax connection (not counting the fourth-order pole sourced by the ghosts).
\subsection{2D Action}
Let us now show that, as expected, the 2D action associated with the solution (\ref{psLax}, \ref{psBarLax}) is the usual pure spinor superstring action. To do this, just plug (\ref{psLax}) and (\ref{psBarLax}) in (\ref{CS+Ghosts}). Let's do this term by term. Using (\ref{deltaId}), the first term yields
\begin{align}
&\frac{1}{2\pi i}\int_{\mathcal{M}}\omega\wedge\str\big(\Lax\wedge\bar d \Lax\big)=
\nonumber\\
&\frac{1}{2\pi i}\bigg\{\int_{\mathcal{M}}d^2\xi d^2w\phi(\xi)\;\str\bigg[\big(J_w^{(0)}-N\big)\partial_{\bar \xi}(\xi^{-4})\hat{N}\bigg]+\nonumber \\
&+\int_{\mathcal{M}}d^2ud^2w\phi(u)\;\str\bigg[\big(J_{\bar w}^{(0)}-\hat{N}\big)\partial_{\bar u}(u^{-4})N\bigg]\bigg\}\nonumber\\
=&\frac{1}{2\pi i}\bigg\{\int_{\mathcal{M}}d^2\xi d^2w\phi(\xi)\;\str\bigg[\big(J_w^{(0)}-N\big)\frac{\pi i}{3}\partial^3_\xi\delta_{\xi,0}\hat{N}\bigg]+\nonumber\\
&+\int_{\mathcal{M}}d^2ud^2w\phi(u)\;\str\bigg[\big(J_{\bar w}^{(0)}-\hat{N}\big)\frac{\pi i}{3}\partial^3_u\delta_{u,0}N\bigg]\bigg\}\nonumber\\
=&-\frac{1}{6}\bigg\{\int_{\mathcal{M}}d^2\xi d^2w\partial^3_\xi\phi(\xi)\;\str\bigg[\big(J_w^{(0)}-N\big)\delta_{\xi,0}\hat{N}\bigg]+\nonumber\\
&+\int_{\mathcal{M}}d^2ud^2w\partial^3_u\phi(u)\;\str\bigg[\big(J_{\bar w}^{(0)}-\hat{N}\big)\delta_{u,0}N\bigg]\bigg\}.
\end{align}
Integrating along $\mathds{CP}^1$ with the delta functions gives
\begin{align}
    &\frac{1}{2\pi i}\int_{\mathcal{M}}\omega\wedge\str\big(\Lax\wedge\bar d \Lax\big)=
    \nonumber\\
    &-\int_{\Sigma} \;d^2w \str\big(J^{(0)}_w\hat N+J^{(0)}_{\bar w}N-2N\hat{N}\big).
\end{align}
The second term can be directly computed to yield
\begin{align}
  &\frac{1}{2\pi i }\sum_k\oint_{\mathcal{C}_k}\omega(u)\int_\Sigma\str\big(\mathcal{L}\wedge\rho^k\cdot J\big) =\nonumber\\
  &=\int d^2 w\;\str\big(J_{\bar w}^{(2)}J_w^{(2)}+\frac{3}{2}J_w^{(3)} J_{\bar w}^{(1)}+\frac{1}{2}J_w^{(1)}J_{\bar w}^{(3)}+\nonumber\\
  &\;\;\;\;\;\;\;\;\;\;\;\;\qquad\qquad\qquad\qquad\qquad+\hat{N}J^{(0)}_w+N J^{(0)}_{\bar w}\big).
\end{align}

The third term only depends on the regular part of the Lax connection and is
\begin{align}
    2\int d^2 w \;\str\big(&y\partial_{\bar w}\lambda+Y\comm{J_{\bar w}^{(0)}-\hat{N}}{\lambda}+
    \nonumber\\+&\hat y\partial_{ w}\hat \lambda+\hat y\comm{J_w^{(0)}-N}{ \hat \lambda}\big)\\
    =\int d^2 w\;\str\big(2&y\nabla_{\bar w}^{(0)} \lambda+2\hat{y}\nabla_{w}^{(0)}\hat \lambda-4N\hat{N}\big),
\end{align}
where 
\begin{align}
\nabla^{(0)}=d+\comm{J^{(0)}}{\cdot}.    
\end{align}
Summing all contributions, the action is proportional to
\begin{align}
   \int d^2 w \;\str\big(\frac{1}{2}J_{\bar w}^{(2)}J_w^{(2)}+\frac{3}{4}J_w^{(3)} J_{\bar w}^{(1)}+
   +\frac{1}{4}J_w^{(1)}J_{\bar w}^{(3)}+\nonumber\\+y\nabla^{(0)}_{\bar w} \lambda+\hat{y}\nabla^{(0)}_w \hat{\lambda}-N\hat N\big
    )
\end{align}
which is the action for the $\ads$ pure spinor superstring.

\subsection{Lagrange multipliers and the BRST charge} \label{LagrangeBRST}

The BRST charge obtained in section \ref{gaugeFixSec} is
\begin{align}
    Q_{\lambda,\hat\lambda}=\oint dw\; \str\big(l\lambda\big)+\oint d\bar w\; \str(\hat l\hat{\lambda}).
    \label{4D_BRST_2}
\end{align}
Since only $l_\alpha$ and $\hat l_{\hat\alpha}$ appear in (\ref{4D_BRST_2}), we can use the other multipliers to impose the associated gauge fixing conditions. Then, the relevant terms in (\ref{CS+Ghosts}) for the computation of ($l,\;\hat l$) are
\begin{align}
    \frac{1}{2\pi i}\int_{\mathcal{M}}\omega \wedge\str\big(\Lax\wedge\bar \partial\Lax\big)-2\int_{\Sigma}d^2 w\str\big(l\Lax_{\bar w}^{-1}+\ \hat l \Lax_{w}^{1}\big) .\label{ContributingForBRST}
\end{align}
Since the multipliers couple to Laurent modes of $\Lax$, they are related to other modes through equations of motion. As $\Lax$ is meromorphic on-shell, let's perform a Laurent expansion on $\Lax$ in the first term of (\ref{ContributingForBRST}). In particular, we are interested in the terms for which either $\Lax^{-1}_{\bar w}$ or $\Lax_w^{1}$ appear. They are
\begin{align}
        &\int_{\mathcal{M}}\omega \wedge\str\big(\Lax\wedge\bar \partial\Lax\big)=\nonumber\\
        &=-4\pi i\int_{\mathcal{M}}d^2u d^2w\;\str\big(\delta_{u,0}\Lax_w^{-3}\Lax_{\bar w}^{-1}-\delta_{u,\infty}\Lax_{\bar w}^{3}\Lax_w^{1}+\ldots\big)\nonumber\\
        &=-4\pi i\int_{\Sigma}d^2w\;\str\big(\Lax_w^{-3}\Lax_{\bar w}^{-1}-\Lax_{\bar w}^{3}\Lax_w^{1}+\ldots\big),
\end{align}
and so varying with respect to $\Lax_{\bar w}^{-1}$ and $\Lax_w^{1}$ in (\ref{ContributingForBRST}) gives 
\begin{align}
    &l=-\Lax_w^{-3}\\
    &\hat l=\Lax_{\bar w}^{3}.
\end{align}
Then, from the solution (\ref{psLax}, \ref{psBarLax}) $\Lax^{-3}_{w}=J_{w}^{(3)}$ and $\Lax^{3}_{\bar w}=J_{\bar w}^{(1)}$. Thus, the BRST charge is
\begin{align}
    Q_{\lambda,\hat\lambda}=-\oint dw\;\str\big( J^{(3)}_{w}\lambda\big)+\oint d\bar w\;\str\big(J_{\bar w}^{(1)}\hat\lambda\big), \label{BRSTfromGF}
\end{align} 
which, up to an irrelevant overall sign, is the correct BRST charge for the pure spinor superstring in $\ads$ (\ref{PureSpinorBRST}).
\acknowledgments
NB would like to thank Kevin Costello and Masahito Yamazaki for useful discussions, and FAPESP grants 2021/14335-0, 2019/21281-4, 2019/24277-8 and CNPq grant 311434/2020-7 for partial financial support. RP 
would like to thank João Gomide, Eggon Viana, and Lucas N.S. Martins for useful discussions,
and FAPESP grant 2022/05236-1 for partial financial
support.
\appendix
\section{Pure spinors and the matrices ($\mathcal{N}$, $\bar{\mathcal{N}}$)} \label{PsAppendix}
A ten-dimensional pure spinor $\lambda^{\alpha}$ is defined to satisfy
\begin{equation}
    \lambda^{\alpha}\gamma^a_{\alpha\beta}\lambda^{\beta}=0 \label{PSconstraint}
\end{equation}
Another useful way to write (\ref{PSconstraint}) is by defining the $\mathfrak{g}$-valued object $\lambda:=\lambda^\alpha Q_\alpha$. Then, from (\ref{QQcomm}), the pure spinor constraint can also be written as 
\begin{equation}
    \acomm{\lambda}{\lambda}=0
\end{equation}
Although not apparent, there are only 5 independent constraints in (\ref{PSconstraint}). This can be seen by breaking the $\mathrm{SO}(10)$ chiral spinor in terms of representations of $\mathrm{U}(5)$ as
\begin{align}
    &\mathbf{16}\to (\mathbf{1},\mathbf{10}, \mathbf{\bar{5}} )\\
    &\lambda^{\alpha}\to (\lambda_+,\lambda_{[ab]},\lambda^a)\;\;\; a=1,\ldots,5.
\end{align} 
In this language, the pure spinor constraint reads
\begin{equation}
    \lambda_+\lambda^a-\frac{1}{8}\epsilon^{abcde}\lambda_{bc}\lambda_{de}=0 \label{U5_PS}
\end{equation}
which proves that there are only 5 independent constraints.
Now, decomposing $(Z\gamma^a Z)$ in terms of $\mathrm{SO}(4,1)\times\mathrm{SO}(5)$ yields
\begin{equation}
    \Phi_I(Z):=(Z\gamma_I Z)\;\;\;\;\Psi_{\tilde I}:=(Z\gamma_{\tilde I} Z)\;\;\; I=0,\ldots 4\;\;\;\;\tilde I=5\ldots{9}
\end{equation}
Since the pure spinor constraint has 5 independent components
\begin{equation}
    \Phi_I(Z)=0\iff \Psi_{\tilde I}(Z)=0 \label{PsEquivalence}
\end{equation}
and $Z^\alpha$ is a pure spinor if $\Phi_I(Z)=0$. Moreover, (\ref{PsEquivalence}) implies that there exists an invertible matrix $M_{\tilde{J}}^I$ such that
\begin{equation}
    \Psi_{\tilde J}(Z)=M_{\tilde{J}}^I(Z)\Phi_I(Z).
\end{equation}
We can then define the matrix $\mathcal{N}_I^a(Z)$ which satisfies
\begin{equation}
    Z\gamma_a Z=\mathcal{N}^I_a(Z)\Phi_I(Z) \label{Ndefinition}
\end{equation}
where $\mathcal{N}_a^I=\delta_a^I$ for $a=0,\ldots,4$ and $\mathcal{N}_a^I=M_a^I$ for $a=5,\ldots,9$. Since $(Z\gamma^aZ)(Z \gamma_a Z)=0$, $\mathcal{N}$ satisfies 
\begin{equation}
    \eta^{ab}\mathcal{N}^I_a\mathcal{N}^J_b=0. \label{nullIdentity}
\end{equation}
This discussion also holds for right-moving spinors $\hat Z^{\hat\alpha}$, which allows us to define $\bar{\mathcal{N}}^I_a(\hat{Z})$ satisfying analogous identities. 
Assuming that the $10\times10$ matrix obtained by joining ($\mathcal{N}_a^I$, $\bar{\mathcal{N}_b^J}$) is invertible, we can decompose a generic $\mathrm{SO}(1,9)$ vector $v_a$ as 
\begin{equation}
    v_a=\mathcal{N}^I_a \upsilon_I+\bar{\mathcal{N}}^I_a\bar{\upsilon}_I
\end{equation}
where ($\upsilon^I$,$\bar\upsilon^I$) is a pair of $\mathrm{SO}(5)$ vectors. In particular, note that due to (\ref{nullIdentity})
\begin{align}
    &\mathcal{N}_a^I v^a=0\implies v_a=\mathcal{N}^I_a \upsilon_I \label{GF_Cons1}\\
    &\bar{\mathcal{N}}^I_a v^a=0\implies v_a=\bar{\mathcal{N}}^I_a\bar{\upsilon}_I\label{GF_Cons2}
\end{align}
\bibliography{main.bib}
\end{document}